\documentclass[twocolumn,prl,twocolumn,superscriptaddress]{revtex4-1}
\usepackage{amsmath,amssymb,mathrsfs}
\usepackage{natbib}
\usepackage{subfigure}
\usepackage{tabularx}
\usepackage{epsfig}
\usepackage{longtable}
\usepackage{amsfonts}
\usepackage{rotating}
\usepackage{subfigure}
\usepackage{amsmath}
\usepackage{comment}

\usepackage{bbold}

\def\be{\begin{equation}}
\def\ee{\end{equation}}
\def\bea{\begin{eqnarray}}
\def\eea{\end{eqnarray}}
\def\nn{\nonumber}

\usepackage{wordlike}

\usepackage[unicode=true,bookmarks=true,bookmarksnumbered=false,bookmarksopen=false,breaklinks=false,pdfborder={0 0 1},backref=false,colorlinks=true]{hyperref}

\hypersetup{linkcolor=magenta,urlcolor=blue,citecolor=blue,pdfstartview={FitH},hyperfootnotes=false,unicode=true}

\begin{document}
\title{Quantum Adiabatic Doping with Incommensurate Optical Lattices}
\author{Jian Lin$^\dag$}
\affiliation{State Key Laboratory of Surface Physics, Institute of Nanoelectronics and Quantum Computing, and Department of Physics, Fudan University, Shanghai 200433, China}
%\affiliation{Collaborative Innovation Center of Advanced Microstructures, Nanjing 210093, China}
\author{Jue Nan$^\dag$ }
\affiliation{Shanghai Branch, National Laboratory for Physical
Sciences at Microscale and Department of Modern Physics,
University of Science and Technology of China, Shanghai 201315, China} 
\affiliation{CAS Center for Excellence and Synergetic Innovation
Center in Quantum Information and Quantum Physics,
University of Science and Technology of China, Hefei, Anhui 230026, China} 
\author{Yuchen Luo}
\affiliation{State Key Laboratory of Surface Physics, Institute of Nanoelectronics and Quantum Computing, and Department of Physics, Fudan University, Shanghai 200433, China}
\author{Xing-Can Yao}
\affiliation{Shanghai Branch, National Laboratory for Physical
Sciences at Microscale and Department of Modern Physics,
University of Science and Technology of China, Shanghai 201315, China} 
\affiliation{CAS Center for Excellence and Synergetic Innovation
Center in Quantum Information and Quantum Physics,
University of Science and Technology of China, Hefei, Anhui 230026, China} 
\author{Xiaopeng Li}
\email{xiaopeng\_li@fudan.edu.cn}
\affiliation{State Key Laboratory of Surface Physics, Institute of Nanoelectronics and Quantum Computing, and Department of Physics, Fudan University, Shanghai 200433, China}
%\affiliation{Department of Physics, Harvard University, Cambridge, Massachusetts 02138, USA}
\affiliation{Collaborative Innovation Center of Advanced Microstructures, Nanjing 210093, China}
%{$^\dag$ These two authors contributed equally to this work.}

\begin{abstract}
{Quantum simulations of Fermi-Hubbard models have been attracting considerable efforts in the optical lattice research, with the ultracold anti-ferromagnetic atomic phase reached at half filling in recent years. An unresolved issue is to dope the system while maintaining the low thermal entropy. Here we propose to achieve the low temperature phase of the doped Fermi-Hubbard model using incommensurate optical lattices through adiabatic quantum evolution. In this theoretical proposal, we find that one major problem about the adiabatic doping is atomic localization in the incommensurate lattice, potentially causing exponential slowing down of the adiabatic procedure. We study both one- and two-dimensional incommensurate optical lattices, and find that the localization prevents efficient adiabatic doping in the strong lattice regime for both cases.  With density matrix renormalization group calculation, we further show that the slowing down problem in one dimension can be circumvented by considering interaction induced many-body delocalization, which is experimentally feasible using Feshbach resonance techniques.  This protocol is expected to be efficient as well in two dimensions where the localization phenomenon is less stable.   
%with density matrix renormalization group, and mean field theories, respectively, and show that the localization, or the exponential-slowing down problem, can be circumvented by considering interaction-induced many-body delocalization, which is  experimentally accessible using Feshbach resonance techniques. 
}

\end{abstract}

\date{\today}

%\pacs{67.85.-d, 03.75.Mn, 05.30.Jp, 05.30.Rt}

\maketitle

%\ncsection{\it Introduction.---}
{\it Introduction.---}
Quantum simulation with ultracold atoms confined in optical lattices have attracted tremendous efforts in the last decade~\cite{bloch2018quantum}. One ultimate goal is to study  quantum many-body physics at low temperature~\cite{1998_Zoller_Jaksch_PRL,2002_Hofstetter_Cirac_PRL,2007_Lewenstein_AP,2008_Bloch_Dalibard_RMP,2015_Lewenstein_RPP,2016_Li_Liu_RPP} whose  simulation on classical computers is subjected to unreachable computational complexity. In particular, quantum simulation of the low-temperature phase diagram of the Fermi-Hubbard model~\cite{2002_Hofstetter_Cirac_PRL,2010_Esslinger_CMP} is of great experimental interest, as it would help understand the strongly correlated physics of direct relevance to making high-temperature superconductors~\cite{2007HighTc}. Owing to the high controllability of the microscopic degrees of freedom in the optical lattice emulator, the question of how the strongly correlated physics emerges can be addressed in an unarguable manner. 

In theory, it is now well-known that Fermi-Hubbard like models could support various exotic quantum phases even in the weakly interacting regime. The combination of Fermi-Hubbard model with synthetic gauge fields leads to quantum Hall like topological states~\cite{2005_Osterloh_PRL,2005_Ruseckas_PRL,2009_Liu_PRL,2011_Dalibard_Gerbier_RMP}; its extension to multi-orbital setting gives rise to rich Fermi-surface nesting effects causing plentiful spontaneous symmetry breaking orders~\cite{2011_Hauke_PRA,2014_Liu_NC,2016_Li_Liu_RPP}; its   incorporation of long rang interactions supports unconventional density waves~\cite{2015_Li_NC}.  These theoretical results obtained  at weak interaction  indicate  the strongly interacting regime which is beyond classical simulation capability may contain even more fascinating physics~\cite{2011_Kivelson_PRB}. To study such  exotic physics demands the experimental optical lattices to enter the extremely low temperature, or more precisely the low entropy regime.  

In the last few years, spectacular progress has been made in optical lattice Fermi-Hubbard emulator---the low-temperature antiferromagnetic phase at half filling has been reached~\cite{2013_Greif_Uehlinger_Science,2015_Hart_Duarte_Nature,2016_Greiner_Nature} with quantum microscope techniques~\cite{2015_Haller_Hudson_NatPhys,2015_Cheuk_Nichols_PRL,2015_Parsons_Huber_PRL,2015_Edge_Anderson_PRA,2015_Omran_Boll_PRL,2016_Greif_Parsons_Science,2016_Cheuk_Nichols_PRL,2016_Parsons_Mazurenko_Science,2016_Boll_Hilker_Science,2016_Cheuk_Nichols_Science,2016_Brown_Mitra_arXiv}.  With the experimental developments, Fermi surface nesting related many-body effects now  become accessible. 
%However, how to maintain low entropy in the  doping process in experiments, which is required to approach the more interesting strange metal, pseudo-gap, and eventually d-wave superconducting phases, remains an outstanding open question. 
However, reaching the low entropy region of the doped Fermi-Hubbard model, in order to emulate strongly correlated electronic quantum physics such as the d-wave superconductivity, is experimentally challenging . 
% as it may cause significant heating on top of the low-entropy half-filled antiferromagnetic phase.  

\begin{figure*}
\begin{center}
\includegraphics[width=.9\linewidth]{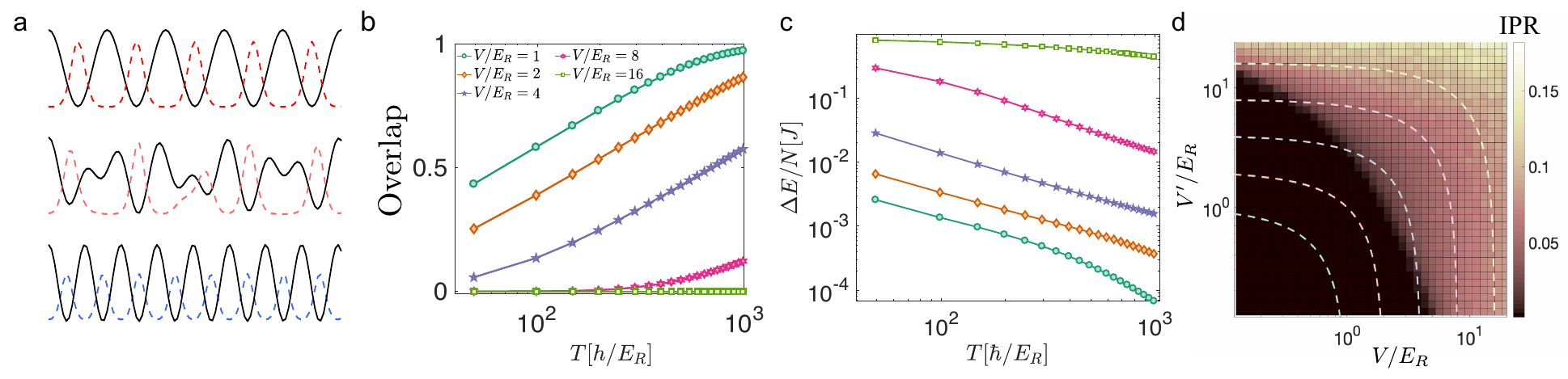}
\end{center}
\vspace{-10pt}
\caption{Quantum adiabatic doping of one-dimensional lattice with free fermions. (a), the schematic illustration of the adiabatic evolution of the lattice. The `dashed' lines in (a) illustrate the evolution of a typical fermion density profile during the quantum adiabatic doping (see more results in Supplementary Material). 
The lattice potential is periodic at both initial and final stage (shown by the top and bottom panels), but the intermediate lattice (the middle panel) is unavoidably incommensurate.  (b), the wave function overlap of the final dynamical state with the ground state of the final Hamiltonian. (c), the excitation energy (per particle) of the final state as compared to the ground state of the final Hamiltonian (see Supplementary Material), where the energy scale $J$ is the single-particle tunneling in the final lattice, and $N$ is the total particle number. 
(c) shares the same legend as shown in (b). 
As we increase the adiabatic time $T$, the overlap in (b) systematically increases and the excitation energy in (c) decreases.  In the adiabatic evolution, we set $V=V'$ 
{(see Eqs.~\eqref{eq:VI} and \eqref{eq:VF}).}  
(d), the inverse participation ratio (IPR)~\cite{vadim,2017_Li_AA_PRB}. The `dashed' lines in (d) correspond to the adiabatic Hamiltonian paths calculated in (b) and (c). The number of periods $L$ ({\it see the main text}) is set  to be $55$ here. 
See the main text for definition of wave function overlap, adiabatic time ($T$), and recoil energy ($E_R$). 
%{\bf The recoil energy $E_R$ is introduced in the main text.} 
} 
\label{fig:onedfree}
\end{figure*}

In this letter, we propose to use incommensurate optical lattices for quantum adiabatic doping. Preparing an initial band insulator state in a periodic optical lattice, the quantum state is adiabatically converted into another lattice having a different period. For a rational filling factor, adiabatic quantum simulation of the Fermi-Hubbard model can be achieved by considering a commensurate superlattice~\cite{2006_Simon_Zoller_PRL,2011_Lubasch_PRL}. To achieve a generic filling in the final optical lattice, the superposed lattice in the intermediate adiabatic process is necessarily incommensurate. 
We study both one- and two-dimensional incommensurate optical lattices, and find that a localization problem occurs when the lattice potential is strong. 
This problem  leads to an exponentially small energy gap, and consequently prevents efficient adiabatic doping for both cases. 
 With density matrix renormalization group calculation, we further show that the slowing down problem in one dimension can be circumvented by considering interaction induced many-body delocalization, which is experimentally feasible using Feshbach resonance techniques.  This protocol is expected to be efficient as well in two dimensions where the localization phenomenon is less stable.   

%By investigating one- and two-dimensional lattices, we show that this problem can be solved in a generic manner by introducing strong atomic interactions through Feshbach resonance techniques, for interaction mediated atomic scattering causes many-body delocalization. These theoretical results are expected to advance quantum simulations of Fermi-Hubbard models in the doped regime. 

{\it Quantum adiabatic doping.---} 
A Fermi-Hubbard optical lattice emulator contains a two-component confined Fermi gas described by the Hamiltonian 
\be 
H = \textstyle \int d^d {\bf x}  \psi_\sigma^\dag \left[-\frac{\hbar^2\vec{\nabla}^2}{2M} + V({\bf x}) -\mu\right] \psi_\sigma 
+ g\psi_\uparrow ^\dag \psi_\downarrow^\dag \psi_\downarrow \psi_\uparrow, 
\ee 
with $M$ the atomic mass, $\mu$ the chemical potential, $V({\bf x})$ the confining optical lattice, and $g$ the interaction strength between the two components. 
In this work, we focus on one- and two-dimensional cases, with $d = 1$, and $2$, respectively. 
For simplicity, we have neglected  harmonic trap potential in the calculation. In experiments, the trap potential should be compensated to keep evaporative cooling in the lattice~\cite{2012_Huse_PRA,2015_Hulet_PRL}.  The initial state we consider is a band insulator in a periodic lattice 
\be 
\textstyle V_I({\bf x}) = V \sum_\alpha \cos (2\pi x_\alpha /\lambda), 
\label{eq:VI}
\ee 
with $x_\alpha = x$ for a one-dimensional lattice, and $\alpha$ labelling the two orthogonal directions for the two dimensional case. 
A relevant energy unit is the single-photon recoil energy $E_R = \frac{\pi^2 \hbar^2}{2M \lambda^2} $. 
With an optical lattice, a band insulator with low thermal entropy is experimentally accessible~\cite{PhysRevLett.120.243201} for the energy gap of the band insulator ground state can be made as large as tens of kHz~\cite{2008_Bloch_Dalibard_RMP}. We then adiabatically convert the system into another lattice having a different period, 
\be 
\textstyle V_F ({\bf x}) = V' \sum_\alpha \cos (2\pi  x_\alpha /\lambda' ), 
\label{eq:VF}
\ee 
with a lattice constant $\lambda ' =\nu \lambda $ 
(see Fig.~\ref{fig:onedfree}(a) for an illustration). 
During the adiabatic evolution, the lattice potential is time-dependent, 
\be 
\textstyle V({\bf x}, t) = \textstyle [ 1- s(t/T)] V_I ({\bf x}) + s( t/T)V_F({\bf x}),  
\label{eq:lattice}
\ee 
with $s(t/T) = t/T$ a standard form  of schedule as in the quantum adiabatic algorithm study~\cite{2018_Lidar_RMP}. Here $T$ is the total adiabatic time. The filling factor (the averaged particle number per degree of freedom) of the final lattice is $\nu^d$, assuming that the time of the adiabatic evolution is much shorter than the atom-loss timescale. Choosing different lattice constant ratio between the initial and final lattices, an arbitrary filling factor can be achieved. 
%The special case with $\nu = 1/2$ corresponds to the half filling which has been studied at Ref.~\onlinecite{2011_Lubasch_PRL}. 
The particle and hole doping are achieved by setting $\nu^d$ greater and smaller than $1/2$, respectively.

{\it Quantum adiabatic evolution of non-interacting fermions.---} 
To demonstrate whether this quantum adiabatic doping works and estimate the required adiabatic time, we first investigate non-interacting fermions in one dimension 
{(see Supplementary Material for the method).} 
In solving the adiabatic evolution (see Eq.~\eqref{eq:lattice}), the tight-binding approximation is not applicable considering the intermediate time region. We thus have to take into account the continuous degrees of freedom of the lattice. In our calculation, the space coordinate is discretized as $x \to j\times a$, with $a$ the grid spacing. A periodic boundary condition is adopted to minimize finite-size effects. Without loss of generality, we set $\nu$ to be the golden ratio 
$ [\sqrt{5}-1]/2$. In the finite size calculation with $L$ number of periods $\lambda_1$, $\nu$ is approximated using Fibonacci sequence $\{F_n\}$, as $\nu \approx F_{n-1}/F_n$, and $L = F_{n-1}$.
The resultant lattice model is 
\bea 
\textstyle H = \sum_j \left \{
\textstyle
 -t \left[ c_{j\sigma}^\dag c_{j+1,\sigma}  + H.c. \right] 
+ V_j c_{j\sigma}^\dag c_{j\sigma} +  Un_{j\uparrow} n_{j\downarrow} \right\},  \nn \\
\label{eq:Hlattice}
\eea
with $t = \frac{\hbar^2}{2Ma^2}$, $V_j = V(ja)$, and $U = g/a$. The interaction term $U$ is $0$ for free fermions. In the numerical calculation, we divide each period of the original lattice potential evenly into $20$ grids. 
We expect the physics presented below to be generic against different choices of irrational filling or boundary conditions (Supplementary Materials). 
%For other choices of irrational filling, we do not expect the physics 

The results are shown in Fig.~\ref{fig:onedfree}. The wave function overlap between the final state of the adiabatic evolution  ($|\Psi_{\rm final}\rangle$) 
and the ground state of the final Hamiltonian ($|\Psi_g\rangle$) is defined as 
\be 
\textstyle {\rm Overlap} = |\langle \Psi_g |\Psi_{\rm final}\rangle |,
\label{eq:overlap}
\ee   
with the result shown in Fig.~\ref{fig:onedfree}(b). 
%shown in Fig.~\ref{fig:onedfree}(b)  overlap of the final quantum state ($|\Psi_{\rm final}\rangle$) with the ground state of the final Hamiltonian ($|\Psi_g\rangle$), i.e., 
%{\bf ${\rm Fidelity} = |\langle \Psi_g |\Psi_{\rm final}\rangle| .$ 
%The infidelity is correspondingly defined as ${\rm Infidelity } =1-{\rm Fidelity}$.  
Fig.~\ref{fig:onedfree}(c) shows the excitation energy in the final state as compared to the ground state. It is evident that the adiabatic quantum evolution is efficient in preparing the final ground state when the lattice is not too deep, say $V' /E_R =  1, 2, 4$. For a deep lattice, e.g. with $V'/E_R = 16$, the adiabatic evolution is found to be no longer efficient---the final state wave function overlap is approximately $0$ and the excitation energy is significant. In Fig.~\ref{fig:onedfree}(d), we show the inverse participation ratio, which is finite (vanishing) in the localized (extended) phase~\cite{vadim, 2017_Li_AA_PRB}. The `dashed' lines in this plot correspond to the parameter-path of the adiabatic evolution. We find that the path enters into the localized regime for the inefficient adiabatic evolution at deeper lattice depths. This shows that the breakdown of the adiabatic preparation corresponds to the atom localization in the intermediate dynamics.  In presence of localization, the level repulsion or the avoided crossing disappears due to the emergent local integrals of motion~\cite{2013_Serbyn_PRL,2014_Huse_MBL_PRB,chandran2015,Ros2015420}, and the minimal energy gap between the ground  and  first excited states becomes exponentially small. The localization thus causes the breakdown of the proposed adiabatic doping process.  Also worth mentioning here is that the localization problem cannot be resolved by upgrading the linear ramp function $t/T$ in Eq.~\eqref{eq:lattice} to nonlinear ones.

\begin{figure}
\begin{center}
\includegraphics[width=\linewidth]{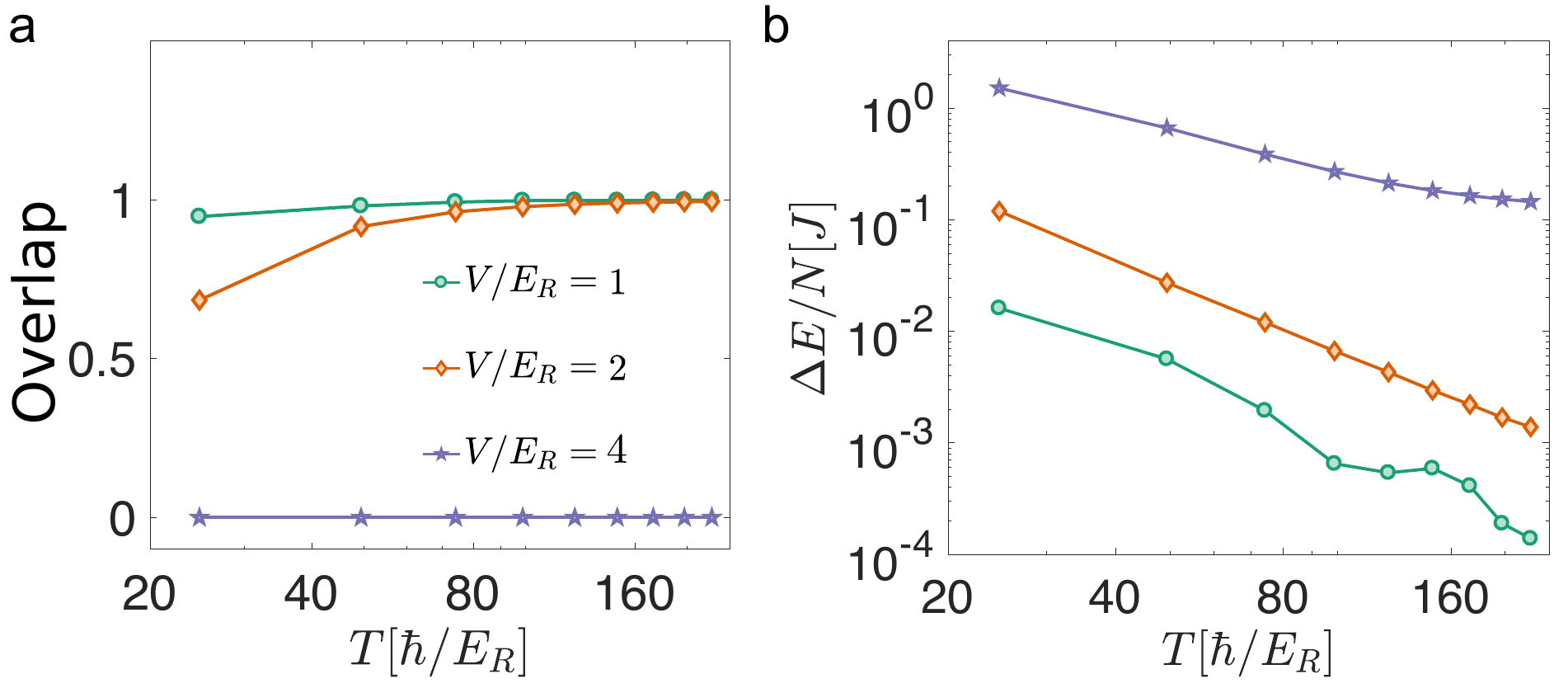}
\end{center}
\vspace{-10pt}
\caption{Quantum adiabatic doping of two-dimensional lattice with free fermions. (a), the final state wave function overlap (see Eq.~\eqref{eq:overlap}) following the two-dimensional adiabatic doping.   (b), the excitation energy (per particle) of the final state. 
{(See Supplementary Material for a detailed description.)} 
Like in the one-dimensional case, as we increase the adiabatic time $T$, we find the systematic increase of the wave function overlap  and the decrease in the excitation energy.  In the adiabatic evolution, we set $V=V'$ (see Eqs.~\eqref{eq:VI} and~\eqref{eq:VF}). (b) shares the same legend as shown in (a). 
In (b), the energy scale $J$ is the single-particle tunneling in the final lattice, and $N$ is the total particle number. 
In the two-dimensional incommensurate lattice, the localization problem that prevents efficient adiabatic doping becomes worse as compared to the one-dimensional case. 
Here we simulate 25 unit-cells (five periods along each dimension) of the original lattice potential. 
%We simulate $25$ unit cells of the original lattice potential in this plot. 
The recoil energy $E_R$ is introduced in the main text. 
} 
\label{fig:2dfree}
\end{figure}

%More relevant to the experimental interest in the Fermi-Hubbard model is the two dimensional case. 
We further consider the two-dimensional case, which is also of great experimental interest~\cite{2013_Greif_Uehlinger_Science,2015_Hart_Duarte_Nature,2016_Greiner_Nature}. 
In Fig.~\ref{fig:2dfree}, we show the results of  wave function overlap and excitation energy following the quantum adiabatic procedure using a two-dimensional incommensurate optical lattice 
{(see Supplementary Material for the method).} 
We still choose $\nu$ to be the golden ratio, so the eventual filling of the two dimensional lattice is $\nu^2=[\sqrt{5}-1]^2/4$. Like in the one-dimensional case, for a lattice not too deep, say with $V/E_R \le 2$, the final state wave function overlap is close to $1$ and the excitation energy is tiny with a reasonable choice of adiabatic time, and the adiabatic doping is thus efficient. With a deeper lattice, the doping process becomes less efficient and eventually fails to prepare the doped ground state due to localization (see IPR in Supplementary Material). 
The localization problem is thus generic to the protocol of quantum adiabatic doping with incommensurate lattices, regardless of the dimensionality. 

{\it Atomic interaction comes to rescue the adiabatic doping.---} 
To rescue the adiabatic doping against localization, one can perform Hamiltonian path optimization through the standard quantum control methods such as GRAPE~\cite{khaneja2005optimal,2011_Machnes_PRA,de2011second}, CRAB~\cite{2011_Caneva_PRA,2011_Doria_PRL}, or Krotov algorithms~\cite{2011_Eitan_PRA,reich2012monotonically}, which is expected to be helpful for experimental implementation of the quantum adiabatic doping. But the problem with that approach is that it is not generically applicable and may require optimization case-by-case for different lattice depths. 

\begin{figure}
\begin{center}
\includegraphics[width=.9\linewidth]{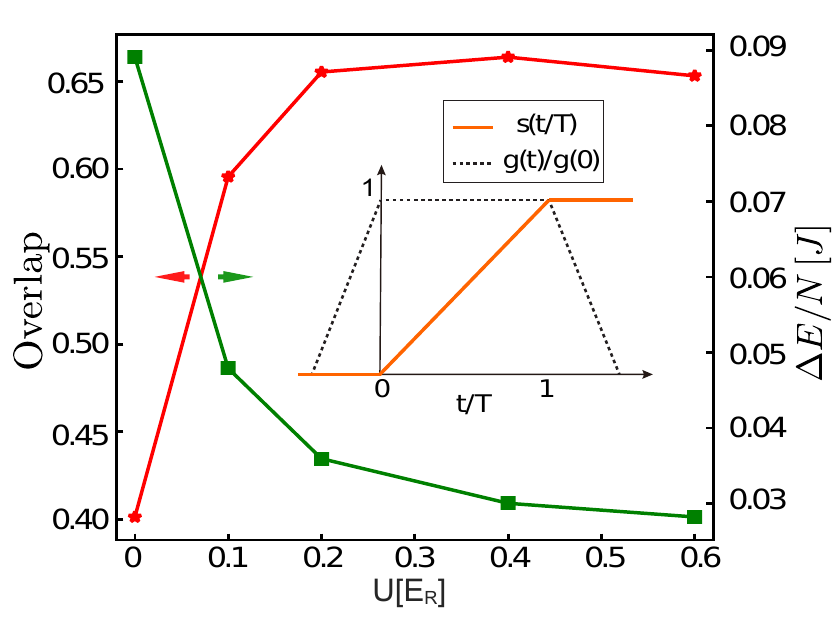}
\end{center}
\vspace{-10pt}
\caption{Performance of quantum adiabatic doping with one-dimensional interacting fermions. We simulate the adiabatic procedure with density matrix renormalization group (DMRG) calculation, taking the Hamiltonian in Eq.~\eqref{eq:Hlattice}. 
The inset shows the time sequence of the adiabatic evolution. The interaction is adiabatically turned on before tuning the lattice potential strengths, then held constant, and adiabatically turned off after the final desired lattice potential is reached.  To show how interaction affects the adiabatic evolution, we carry out DMRG calculation in the regime $t/T\in[0,1]$, where the interaction is held constant.  
%In the calculation, we choose the number of periods $L = 13$, with each period divided into $20$ pieces. 
The lattice potential strengths are $V = V' = 8E_R$ for  which the intermediate regime for the adiabatic evolution is strongly localized in absence of interaction. The adiabatic time is set to be $T = 200$ in the unit of $\hbar/E_R$. It is evident that introducing interaction effects dramatically improve the final state wave function overlap (see Eq.~\eqref{eq:overlap}) shown by the `$\bigstar$'s and the excitation energy  by the `$\blacksquare$'s  following the adiabatic doping, and that increasing interaction strength systematically improves the performance of the procedure.  
Here the energy scale $J$ is the single-particle tunneling in the final lattice, and $N$ is the total particle number. 
See Supplementary Material for details of the method. 
} 
\label{fig:1ddoping}
\end{figure}

A generic approach to solve the localization slowing down problem is offered by the theoretical study of many-body localization---interaction effects tend to generically destabilize localization~\cite{basko2006problem,blochmbl,DemarcoDisorder,2015_Monroe_MBL_NatPhys,2016_Choi_Bloch_MBL_Science}.  We thus introduce an auxiliary Hamiltonian beyond the standard adiabatic quantum computing algorithm. It has been shown in the context of quantum algorithms that this approach could lead to exponential speedup~\cite{van2001powerful,aharonov2008adiabatic,Yu_2018}. The auxiliary Hamiltonian we apply here to optical lattice system is the atomic interaction 
$
\textstyle H_{\rm AU} =  g (t) \int d^d{\bf x} \psi_\uparrow ^\dag \psi_\downarrow^\dag \psi_\downarrow \psi_\uparrow. 
$
The time sequence we consider in this work is shown in the inset of Fig.~\ref{fig:1ddoping}. We first prepare a  noninteracting band insulator in the initial lattice.  Then we turn on the interaction via Feshbach resonance techniques, perform the adiabatic lattice conversion, and then switch off the interaction. 
The interaction strength during the adiabatic lattice conversion has to be large enough to disable the localization. 
For experiments targeting doped Fermi Hubbard model ground state with a particular interaction, one can eventually adiabatically tune the final interaction to that strength. 
We remark here that this work here is to show interaction effects could enable efficient adiabatic doping, 
rather than to optimize the time sequence.    
%How to optimize the time sequence in presence of interaction is left for future study. 
%with the time-dependent  interaction strength that satisfies the boundary condition $g(0) = g(1) = 0$. We choose 
%\be 
%g(t/T) = g_{\rm max} \sin\left(\frac{\pi t}{T}\right).
%\ee 
%This time-dependence can be implemented in cold atoms with Feshbach resonance techniques. 
%We emphasize that the specific form of the interaction is not crucial, becuase the role of this added interaction is to destabilize the localization effects. In this work, we focus on the case with the intial and final Hamiltonian to be non-interacting, but the proposed protocol is expected to perform even better for in interacting case for the localization problem is further destabilized.  

{\it Accelerating the one-dimensional adiabatic doping.---} 
To explicitly demonstrate the feasibility of the adiabatic doping with interaction, we simulate the quantum dynamics of the one dimensional case with density matrix renormalization group (DMRG). 
Our DMRG program is developed according to the method in 
Ref.~\cite{schollwock2011density} using matrix product stats. 
For numerical implementation convenience, we choose an open boundary condition in the DMRG calculation. 
%(See Supplementary Material for details of the method.)
Fig.~\ref{fig:1ddoping} shows the the simulated results with $V/E_R =V'/E_R= 8$, 
{corresponding to one parameter choice in Fig.~\ref{fig:onedfree} where localization occurs in absence of interaction. 
In the DMRG calculation, each period of the lattice potential is still divided into $20$ pieces as in the noninteracting case, but the number of periods simulated is reduced to  $L =13$, due to expensive numerical cost.} 
The dynamics is calculated using the second-order Trotterization with 26000 evolution steps for the adiabatic time of $T = 200\hbar/E_R$. 
We choose the bond dimension  to be $150$, for which the results of final state wave function overlap and excitation energy have converged 
(see Supplementary Material).
In the regime  where localization causes the breakdown of the adiabatic doping procedure, we find that interactions could help significantly enhance the final state wave function overlap and reduce the excitation energy. The performance of the adiabatic procedure can be further improved by increasing the time of adiabatic evolution. After the many-body localization in the adiabatic procedure is suppressed, the required adiabatic time to reach certain level of excitation energy is expected to scale polynomially with the system size for the many-body level repulsion of a thermal quantum system.

{\it Discussion of the two-dimensional case.---} 
For the two-dimensional case, although we could not afford to carry out the calculation due to the computational challenge---this challenge is the precise reason why quantum simulation of Fermi-Hubbard model is needed, we anticipate the interaction also solves the localization problem in two dimensions. In the study of many-body localization~\cite{basko2006problem}, it is expected that the localization phenomenon is less stable in higher dimensions according to the thermal bubble argument~\cite{2016_Choi_Bloch_MBL_Science,2017_Roeck_MBL_PRB,2018_Altman_arXiv,wahl2019signatures}.  The protocol of adiabatic doping combined with auxiliary interaction is thus expected to hold as well for a two-dimensional lattice. 

{\it Experimental timescales.---} 
Commonly used atoms to simulate Fermi-Hubbard model in optical lattice experiments are $^{40}$K and $^6$Li. Considering these atoms confined in an optical lattice formed by laser beams of wavelength   $1064$ nm, the recoil energies are $E_R /\hbar \approx 2\pi  \times 4$ kHz and $2\pi \times 29$ kHz, respectively. The required adiabatic time for the excitation energy per atom to drop down to one percent of single-particle tunneling is within $10$ms according to our calculation. 

{\it Conclusion.---} 
To conclude, in quantum simulations of Fermi-Hubbard model with optical lattices, the adiabatic doping can be achieved by using incommensurate lattices.  For noninteracting fermions, the localization problem makes the adiabatic doping inefficient in both one- and two-dimensional lattices in the strong potential region. As a generic recipe to solve this localization problem, we show that this problem can be solved in a generic manner by introducing strong atomic interactions through Feshbach resonance techniques, for interaction mediated atomic scattering causes many-body delocalization. This is confirmed by DMRG calculation for the one-dimensional system and is expected to hold for the two-dimensional case as well. These theoretical results may improve optical lattice quantum simulations of Fermi-Hubbard models in the doped regime of great  importance to modeling high temperature superconductors or strongly correlated electrons in general.

{\it Acknowledgement.---} 
We acknowledge helpful discussion with Peter Zoller, Immanuel Bloch, Ana Maria Rey, Andrew Daley, Youjin Deng, and Markus Greiner. 
%This work is supported by National Program on Key Basic Research Project of China under Grant No. 2017YFA0304204, 2018YFA0306501, National Natural Science Foundation of China under Grants No. 11774067, 11874340. 
This work is supported by National Program on Key Basic Research Project of China under Grant No. 2017YFA0304204, 2018YFA0306501, National Natural Science Foundation of China under Grants No. 11774067, 11934002, 11874340. 
Calculations were performed based on the ITensor Library~\cite{ITensor}.
X.L would like to thank Department of Physics at Harvard University for hospitality during the completion of this work. 

$^\dag$Jian Lin and Jue Nan contributed equally to this work.

\bibliography{references}
\bibliographystyle{apsrev4-1}

\begin{widetext}
\newpage

\begin{center} 
\Huge{\bf Supplementary Material}
\end{center} 
\renewcommand{\theequation}{S\arabic{equation}}
\renewcommand{\thesection}{S-\arabic{section}}
\renewcommand{\thefigure}{S\arabic{figure}}
\renewcommand{\bibnumfmt}[1]{[#1]}
\renewcommand{\citenumfont}[1]{#1}
\setcounter{equation}{0}
\setcounter{figure}{0}

%\begin{center} 

\section{Numerical method to simulate a many-body state of free fermions}
%\end{center} 
\label{sec:freefermion} 
In this section, the details of our method in simulating non-interacting fermions are provided. The method is described by assuming a lattice Hamiltonian, with no loss of generality, because the Hamiltonian in continuum can be mapped to a lattice Hamiltonian through discretization---the problem of our interest here has no fermion doubling issue. 

Considering a lattice model, with its degrees of freedom described by creation/annihilation operators ($c_x$/$c_x^\dag$)---the subscript $x$ is the lattice-site index. For non-interacting fermions, the Hamiltonian takes a form of 
\be 
H(t) = K + V(t), 
\ee 
where  $K = \sum_{x x'} {\cal K} _{x x'} c_x ^\dag c_{x '} $ represents the time independent kinetic  tunnelling, and $V(t) = \sum_x {\cal V}_{xx'} (t) c_x ^\dag c_{x'} $ the time dependent potential. The matrix ${\cal V}$ is diagonal.
The time-evolving many-body state is described as 
\be 
|\Psi (t) \rangle = \psi_1 ^\dag (t) \psi_2 ^\dag (t ) \ldots \psi_N ^\dag  (t) | 0\rangle, 
\ee  
where $N$ is the total particle number of fermions, and the operators $\psi_{1, 2, \ldots N} $ are related to lattice operators by 
\be 
\psi^\dag _j = \sum_x U_{x j} (t) c_x^\dag  .
\ee
The $U$ matrix contains the single-particle wave function of each occupied orbital. 
From the many-body schr\"odinger equation, $i\partial_t |\Psi (t) \rangle = H(t) | \Psi (t) \rangle$, the $U$ matrix satisfies the single-particle schr\"odinger equation, 
$$i\partial_t U_{x j} (t) = \sum_{x'} \left[ {\cal K}_{xx'} + {\cal V}_{x x'} (t) \right] U_{x' j} (t) .$$

\begin{figure}[htp]
\begin{center}
\includegraphics[width=160mm]{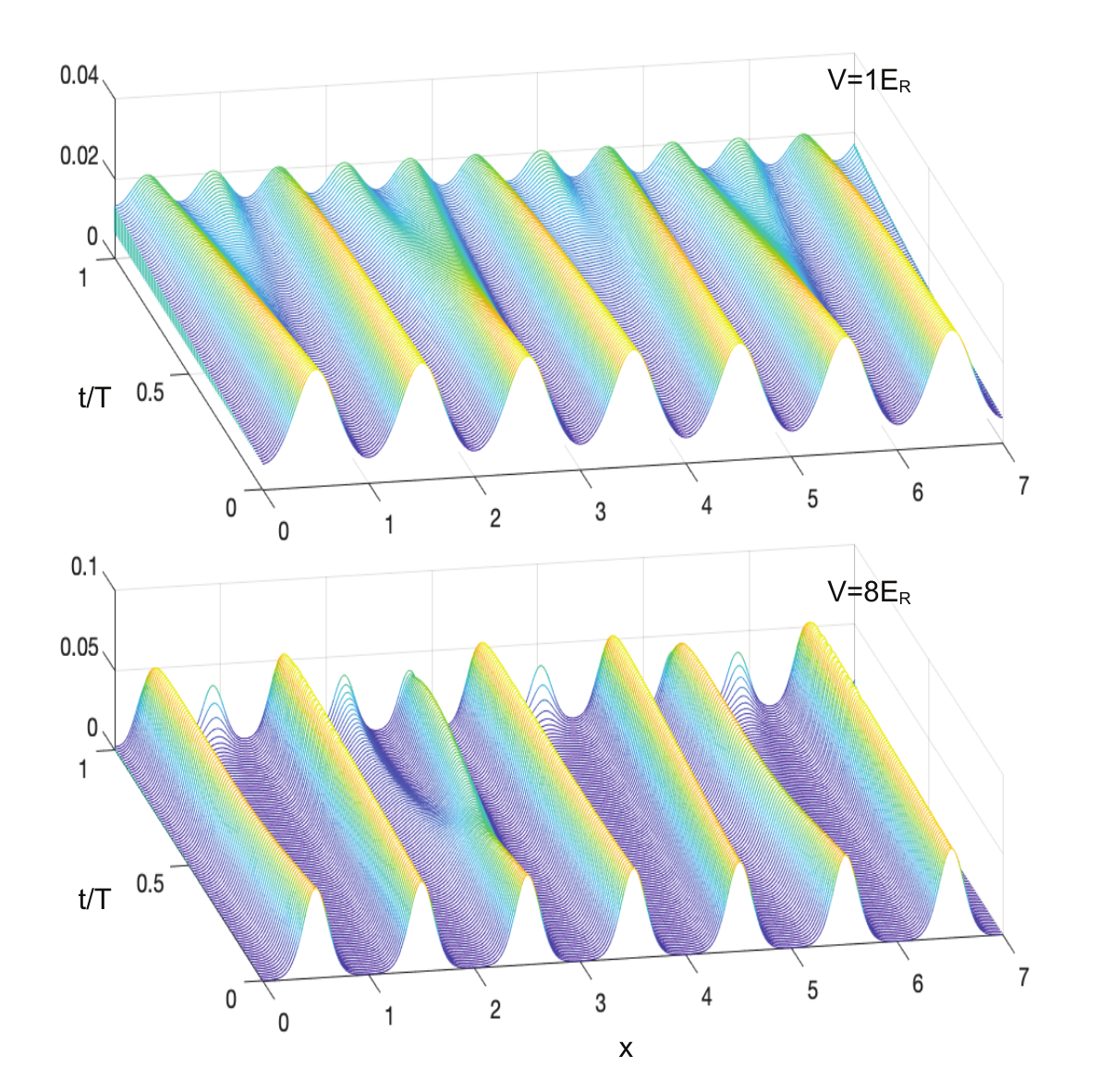}
\end{center}
\caption{The time evolution of density profile in quantum adiabatic doping of the 1d lattice. In the delocalized case ($V = 1E_R$), the density profile follows the instantaneous ground state of the system. In the localized case ($V = 8E_R$), it has a large memory retention of the initial state. The total adiabatic time is set to be $1000\hbar/E_R$ in this plot.  } 
\label{fig:density}
\end{figure}

\begin{figure}[htp]
\begin{center}
\includegraphics[width=100mm]{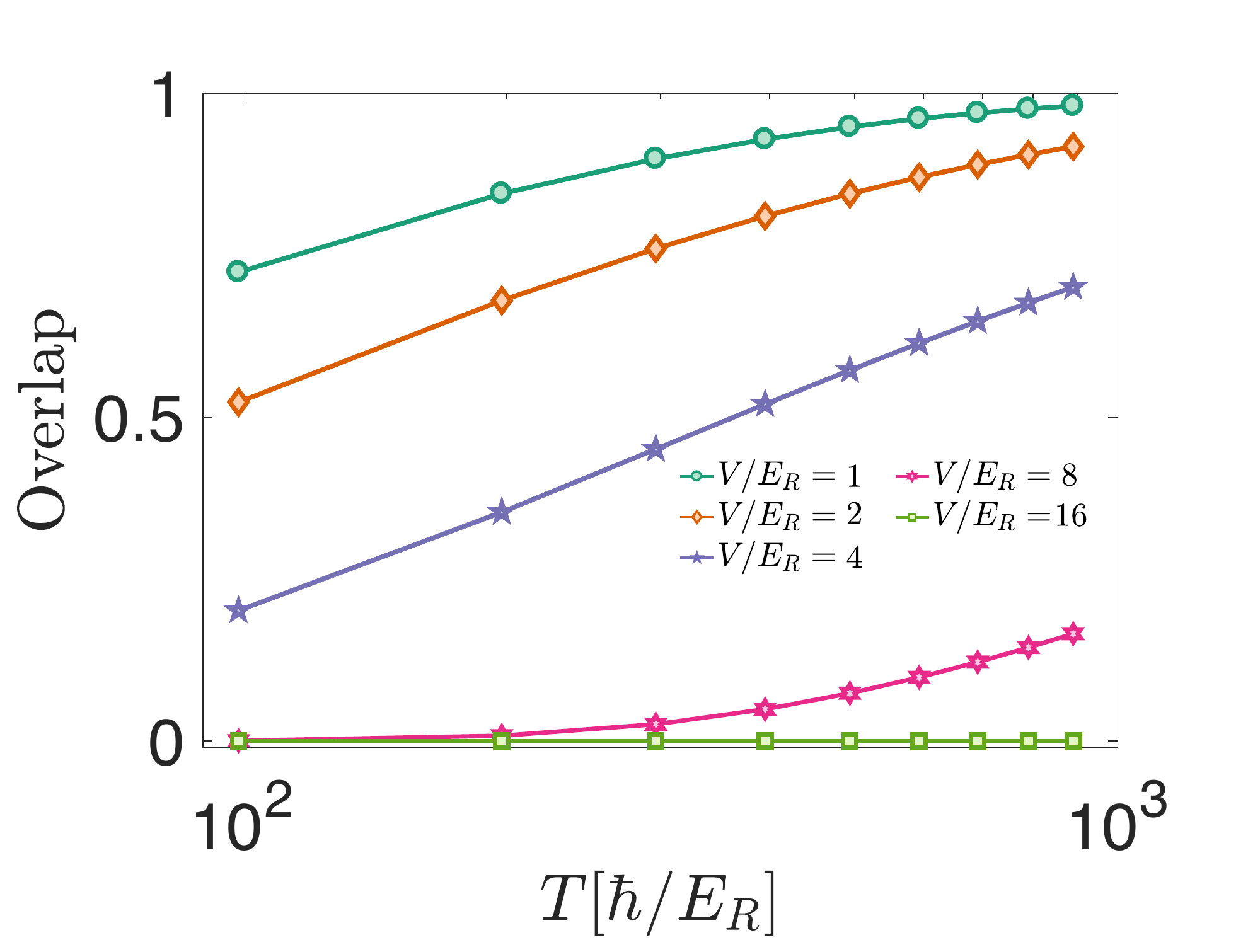}
\end{center}
\caption{The resultant wave function overlap in the quantum adiabatic doping procedure with the open boundary condition. In this plot, the parameter used is the same as in Fig.~1(b) in main text. The only difference here is we use an open boundary condition. By comparison, it is evident that the qualitative feature of the fidelity  remains the same as in Fig.~1 where a periodic boundary condition is used, despite of their quantitative difference due to finite size effects.  } 
\label{fig:OBC}
\end{figure}

\begin{figure}[htp]
\begin{center}
\includegraphics[width=110mm]{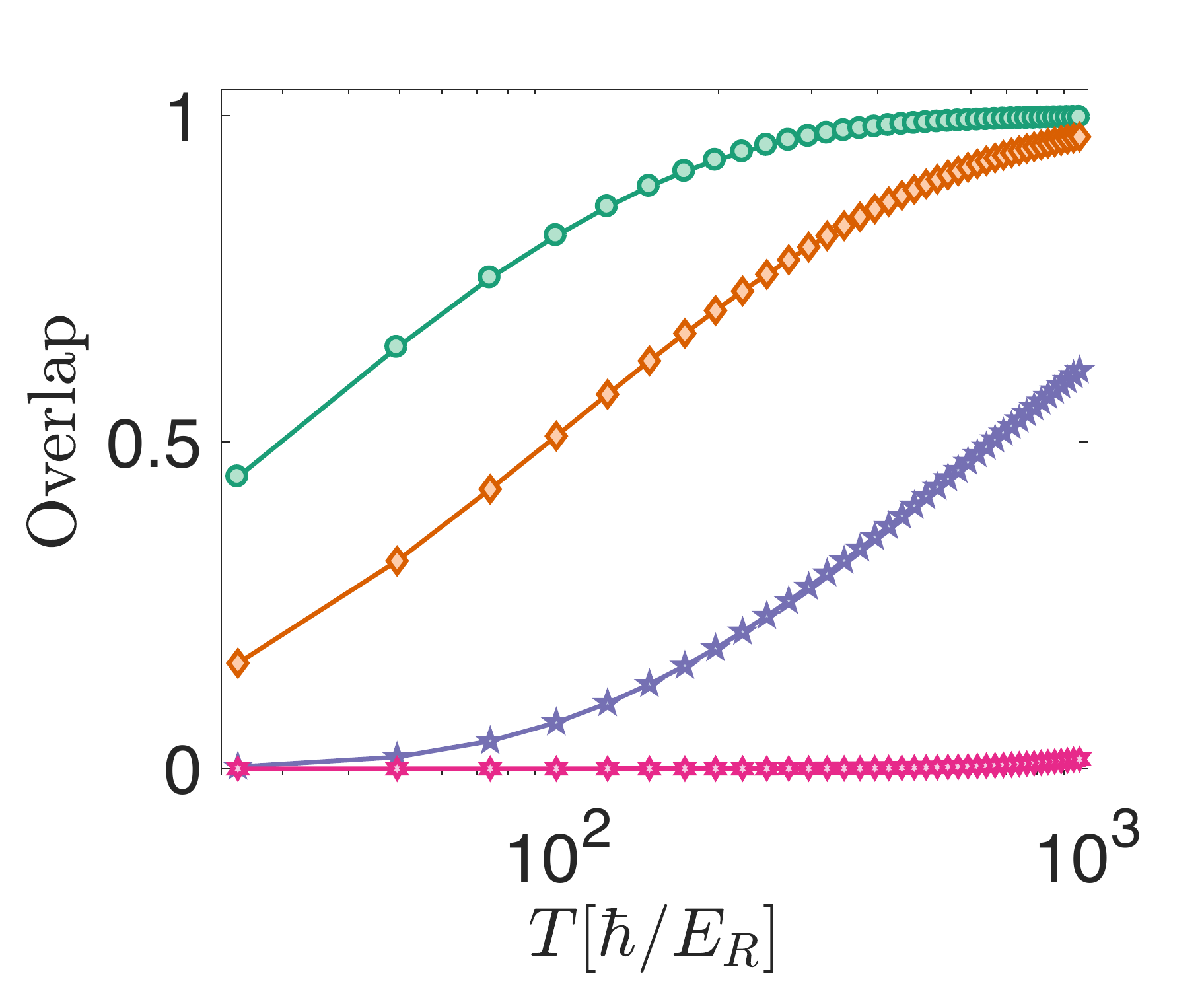}
\end{center}
\caption{The resultant wave function overlap in the quantum adiabatic doping procedure with a filling factor of $\nu = 1/\sqrt{2}$. In this plot, the parameter used is the same as in Fig.~1(b). The only difference here is we choose a filling factor of $\nu = 1\sqrt{2}$. The results shown here is qualitatively similar to Fig.~1(b), where the filling factor is $(\sqrt{5}-1)/2$. This figure uses the same legend as given in Fig.~\ref{fig:OBC}.   
} 
\label{fig:sqrt2filling}
\end{figure}

\begin{figure}
\begin{center}
\includegraphics[width=100mm]{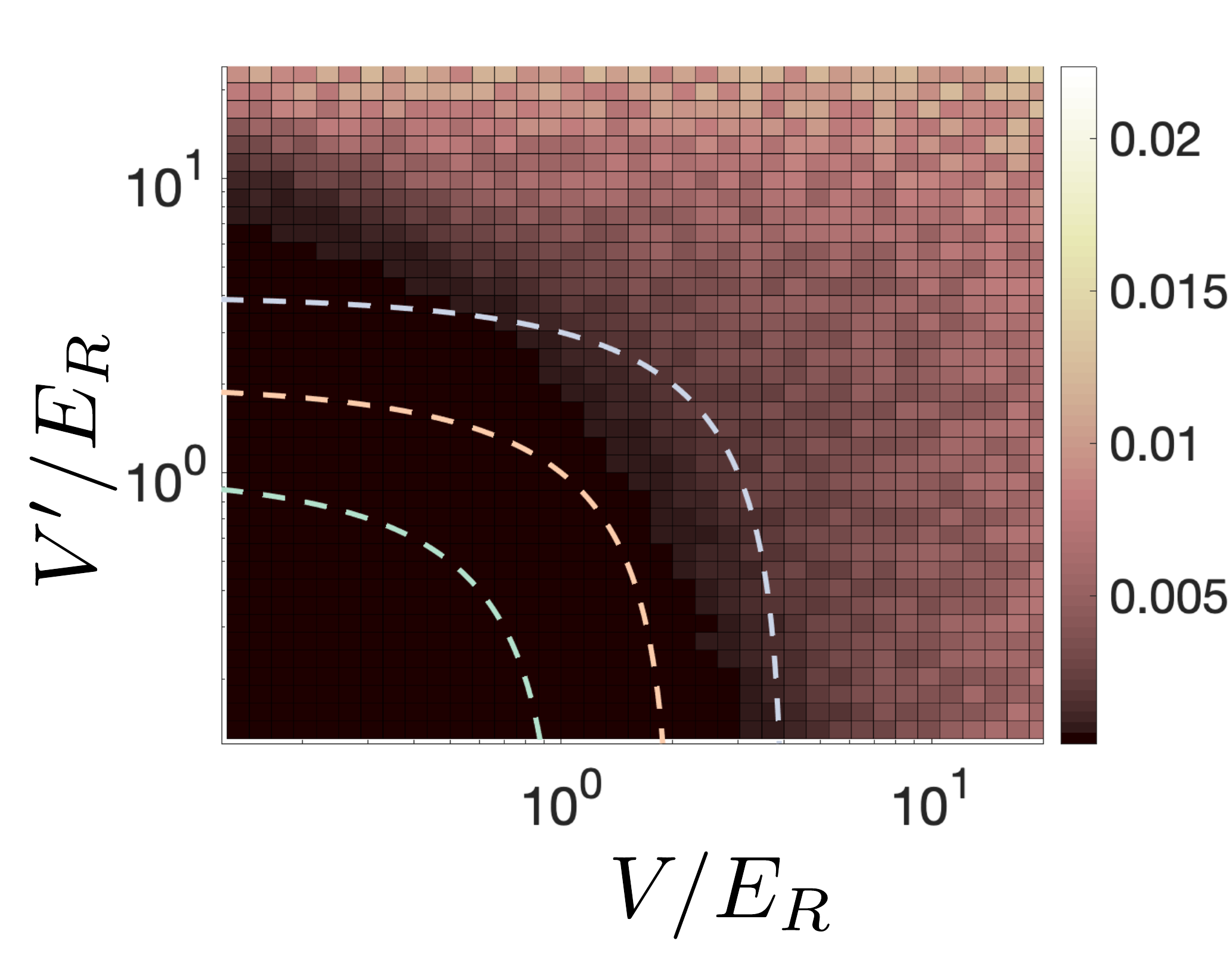}
\end{center}
\caption{The inverse participation ratio for the 2d lattice. The color in this plot index the inverse participation ratio (IPR). The `dashed' lines correspond to the adiabatic Hamiltonian path used in Fig.~2  (main text).   As compared to the 1d case, the dark region which corresponds to the delocalized phase becomes smaller. The finite-size artifact of the 2d case is larger than the 1d case in Fig.1(d) (main text), because the linear size we simulate in 2d is much smaller than 1d.} 
\label{fig:IPR2d}
\end{figure}

In simulating the dynamics, we solve the time-dependent problem numerically by second-order Trotterization, 
\be 
U (t+ \delta t) = e^{-i{\cal K} \delta t/2} e^{-i {\cal V} (t+\delta t/2) \delta t} e^{-i {\cal K} \delta t/2} U(t) +{\cal O} (\delta t ^3) . 
\ee 
Note that the matrix ${\cal V}$ is diagonal, so the multiplication of  $e^{-i{\cal V} \delta t}$ on a vector can be implemented as multiplication of each element by a phase, which is numerically very efficient. The matrix ${\cal K} $ is non-diagonal, but can be diagonalized by a Fourier transformation. Then the multiplication of $e^{-i{\cal K} \delta t}$ can be efficiently carried out  by using Fast-Fourier-Transformation. 
Regarding the initial condition, at time $t = 0$, the initial state in our adiabatic doping procedure is the instantaneous ground state of the Hamiltonian at time $t =0$. 

The  energy of the dynamical state is defined with respect to the instantaneous Hamiltonian, which can be calculated according to 
\be 
E(t) = \sum_{j=1}^N \sum_ {x x'} \left[ {\cal T}_{x x'} + {\cal V} _{x x'} (t)  \right] U_{x j} ^* (t) U_{x'j} (t) . 
\ee 
The excitation energy of the adiabatic procedure  is defined to be energy of the final state subtracted by the ground state energy of the final Hamiltonian. 
We also calculate the wave function overlap between the final state of the dynamical evolution and the ground state of the final Hamiltonian, defined as 
\be 
{\rm Overlap} = |\langle \Psi_g |\Psi(t=T)\rangle| . 
\label{eq:Fidelity}
\ee
With the wave function of the occupied orbitals for $|\Psi_g\rangle$ stored as $U_{x j} ^{(g)}$, the wave function overlap reads as 
\be 
{\rm Overlap} = |{\rm Det} [U^\dag  U^{(g)} ]|. 
\ee 
For the spinful case, the two spin components can be further incorporated in a straightforward way.

\begin{figure}
\begin{center}
\includegraphics[width=160mm]{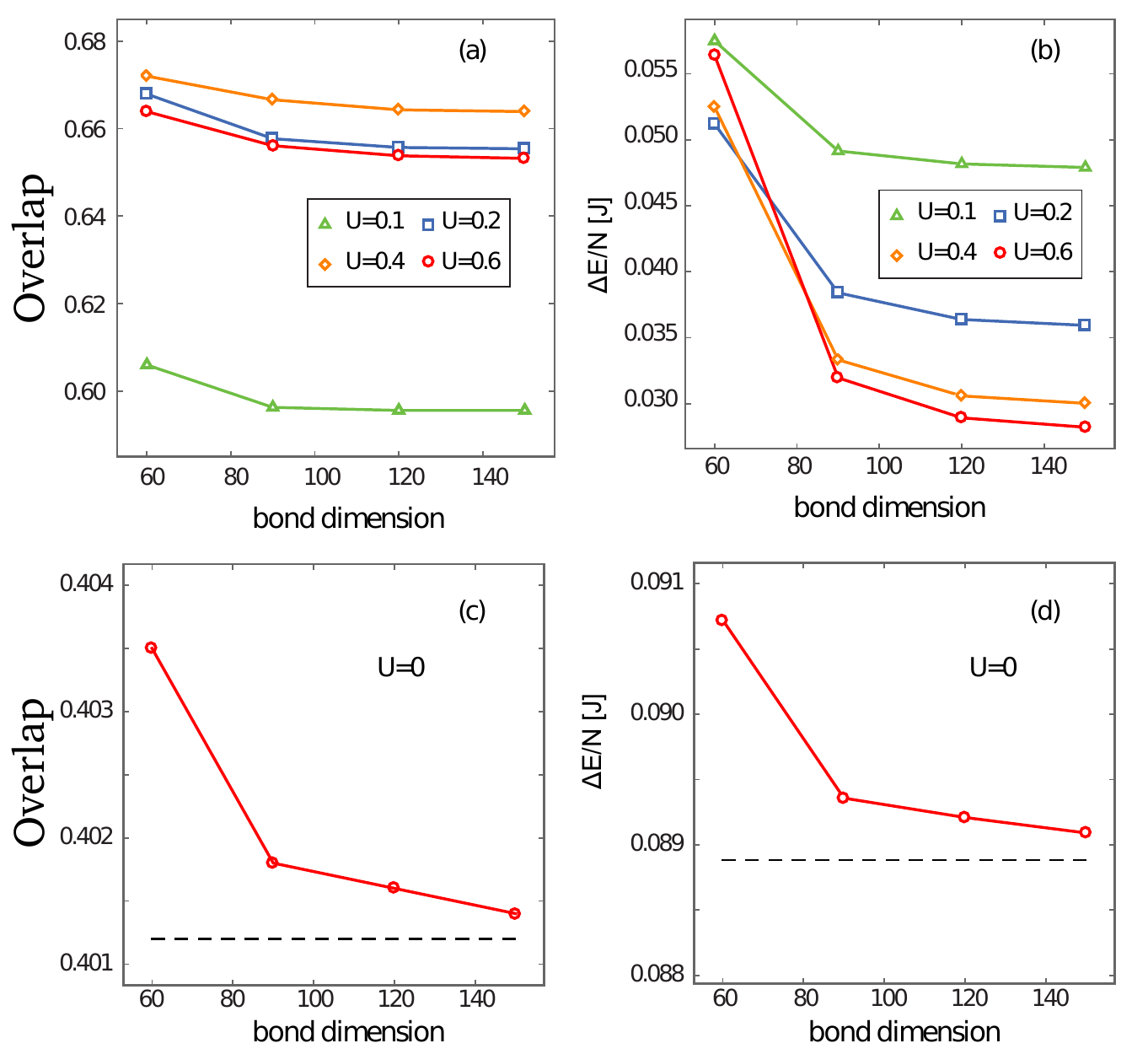}
\end{center}
\caption{
The convergence of the results of wave function overlap (Eq.~\eqref{eq:Fidelity}) and excitation energy in DMRG simulation. Here we choose the lattice depths  $V = V'=8E_R$ 
({\it see Eq. (2,3) in the main text}), the adiabatic time $T = 200 \hbar/E_R$, and $26000$ evolution Trotter-steps. 
(a), the wave function overlap as a function of bond dimension $D$ with different interaction strengths. The relative error of the results between $D=120$ and $D=150$ is within $0.1\%$.   
(b), the excitation energy (per particle) as a function of $D$. The relative error for the results between $D=120$ and $D=150$ is within $2.5\%$. In (b), $N$ is the total particle number, and $J$ is the single-particle tunneling amplitude in the final lattice. 
Both the wave function overlap and the excitation energy have converged at large bond dimension.
(c,d), the comparison of DMRG results (red `solid' curve) with exact simulation (`dashed' line) for free fermions. We find quantitative agreement between the converged DMRG results and exact ones. } 
\label{fig:DMRG}
\end{figure}

\section{Evolution of density profile in the quantum adiabatic doping} 
In our simulation of free fermions in the quantum adiabatic doping, besides the final state fidelity and excitation energy, we also calculate the evolution of the density profile. The results are shown in Fig.~\ref{fig:density}. When the lattice is shallow ($V = E_R$), the density profile follows the instantaneous ground state evolution, and the confined atoms slowly migrate from original lattice sites to the final lattice sites, and eventually becomes evenly distributed in the final lattice. In contrast, for a strong lattice ($V = 8E_R$), the quantum tunneling from the original sites to the final sites is not efficient enough to make an even density profile when the adiabatic doping finishes. 

\section{Robustness against different doping fraction and boundary conditions}
In optical lattice experiments, we typically do not have periodic boundary condition as used for the results of free fermions in the main text. To confirm our proposing  quantum adiabatic doping is adaptable, we check its robustness against the choice of boundary conditions.  From the results shown in Fig.~\ref{fig:OBC}, it is evident that for free fermions the quantum adiabatic doping remains efficient when the lattice is shallow, and becomes inefficient  in a strong lattice. The behavior of the fidelity is qualitatively the same as the case of periodic boundary condition as shown in Fig.~1 (see main text). We thus expect the physics presented in this work to be largely independent of the choice of boundary conditions. 

To confirm the quantum adiabatic doping works for generic fillings, we provide the results for the filling factor $\nu = \sqrt{2}$, which can be asymptotically approached by using the Pell's number sequence. Despite the quantitative difference form the the choice of golden ratio, the generic feature---the efficiency (breakdown) at weak (strong) lattice---remains (see Fig.~\ref{fig:sqrt2filling}). We expect the physics in our quantum adiabatic doping to be generic for different filing factors.

\section{Inverse participation ratio for the 2d lattice} 
Here we provide results of inverse participation ratio of the 2d lattice in the quantum adiabatic doping procedure (see Fig.~\ref{fig:IPR2d}).  It is evident that the breakdown of the quantum adiabatic doping for the 2d case at strong lattice is also due to localization physics as in the 1d case. 
We thus expect the interaction restores the efficiency of the quantum adiabatic doping in two dimensions. In the study of many-body localization~\cite{basko2006problem}, it is expected that the localization phenomenon is less stable in higher dimensions according to the thermal bubble argument~\cite{2016_Choi_Bloch_MBL_Science,2017_Roeck_MBL_PRB,2018_Altman_arXiv,wahl2019signatures}. 

\section{Details of density-matrix-renormalization-group simulation}

In this section, we provide the details of simulating interacting fermions with the density-matrix-renormalization-group (DMRG) method. 
%For the ground state of the one dimension Fermi-Hubbard model with interaction, the standard two-site procedure of DMRG is applied and the truncation-error $\epsilon$ in singular value decomposition is limit to $10^{-15}$. Such choice of $\epsilon$ leads to the energy uncertain $\sqrt{\langle H^2 \rangle - \langle H \rangle^2 } \le 2 \times 10^{-4}$,  which is sufficiently small to estimate the fidelity and excitation energy for adiabatic doping. 
In the adiabatic evolution, we represent the dynamical state with a matrix product state of a  finite bond dimension $D$, and use the standard two-site procedure in simulating the quantum dynamics~\cite{schollwock2011density}. 
% with the bond dimension $D$ of the MPS fixed to $150$. 
The inital state of the adiabatic evolution is the band-insulator ground state of the initial lattice for the adiabatic evolution, which is also represented by a matrix-product state in the calculation. 
To obtain the wave function overlap (see Eq.~\eqref{eq:Fidelity}) and excitation energy of the final quantum state at the end of the adiabatic evolution, the ground state of the final Hamiltonian is also calculated by the DMRG method. 
%which can be efficiently carried out in the framework of matrix-product-state.  
The excitation energy is defined to be energy of the final state of the evolution subtracted by the ground state energy of the final Hamiltonian. 
The results of wave function overlap and excitation energy are shown  in Fig.~\ref{fig:DMRG}. 
We see convergence of these results with the bond dimension $D$ increased from $60$ upto $150$. 
The relative error in excitation energy comparing $D=120$ and $D=150$ is within $2.5\%$ and for the wave function overlap it is within $0.1 \%$. 
We also benchmark our DMRG simulation using the free fermion model with the exact results from the method described in the last section. With the bond dimension $D$ increased to $150$, we find quantitative agreement between the DMRG and the exact results (see Fig.~\ref{fig:DMRG}(c,d)), which justifies the correctness of the converged DMRG results.  

\end{widetext}

\end{document}